\documentclass[twocolumn,11pt]{aastex62}
\usepackage{amsmath}
\usepackage{graphicx}
\usepackage{multirow}
\usepackage{CJK}

\newcommand{\beq}{\begin{equation}}
\newcommand{\eeq}{\end{equation}}
\newcommand{\bea}{\begin{eqnarray}}
\newcommand{\eea}{\end{eqnarray}}

\shorttitle{Tidal stripping of white dwarf} 
\shortauthors{Shen}

\begin{document}

\title{Fast, ultra-luminous X-ray bursts from tidal stripping of White Dwarfs by Intermediate-mass Black Holes}

\begin{CJK*}{UTF8}{gbsn}

\email{shenrf3@mail.sysu.edu.cn}
\author{Rong-Feng Shen (申荣锋)}
\affil{School of Physics \& Astronomy and Institute of Astronomy \& Space Science, Sun Yat-Sen University, Guangzhou, P. R. China}

\begin{abstract}

Two X-ray sources were recently discovered by Irwin et al. in compact companions to elliptical galaxies to show ultra-luminous flares with fast rise ($\sim$ minute) and decay ($\sim$ hour), and with a peak luminosity $\sim 10^{40-41}$ erg s$^{-1}$. Together with two other sources found earlier, they constitute a new type of fast transients which cannot be attributed to neutron stars but might be due to intermediate-mass black holes (IMBHs; $10^{2-4} M_{\odot}$). The flaring behavior is recurrent for at least two sources. If the flare represents a short period of accretion onto an IMBH during the periastron passage of a donor star on an eccentric (i.e., repeating) or parabolic (non-repeating) orbit, we argue that the flare's rise time corresponds to the duration during which the donor's tidally stripped mass joins a residual disk at the pericenter. This duration is in turn equal to three other time scales: the duration of stripping, the sound crossing time of the donor, and the circular orbit time at the pericenter radius. Only a white dwarf can have a sound crossing time as short as one minute. Therefore, the donor must be a white dwarf and it was stripped of $\sim 10^{-10}~M_{\odot}$ upon each passage at several to tens of Schwarzschild radii from the IMBH. The flux decay corresponds to the viscous drainage of the supplied mass toward the hole. Aided with long-term X-ray monitoring, this type of fast transients would be an ideal target for next-generation gravitational wave detectors.      

\end{abstract}

 
\section{Introduction}	    \label{sec:intro}

There are much evidence for the existence of stellar-mass ($\sim 10 M_{\odot}$) and supermassive ($\sim 10^6 - 10^9 M_{\odot}$) black holes (BHs), but still no firm evidence for existence of intermediate-mass black holes (IMBHs;$\sim 10^{3-4} M_{\odot}$) \citep{PZ02, baumgardt03,tremou18}, which fill a gap of the mass range in between. Nevertheless, the search and identification of them has great impact on understanding of the seeds and growth history of SMBHs \citep{volonteri03,greene12}. 

Recently, \citet{irwin16} found two luminous fast flaring sources in nearby galaxies from a search in archival Chandra data. One source is located in a globular cluster in the galaxy NGC4636. It brightens within 22 seconds by a factor of 100 to reach a peak luminosity of $9\times10^{40}$ erg s$^{-1}$, then decays in 1,400 s. The persistent emission before and after the flare is at $8\times10^{38}$ erg s$^{-1}$. 

The second source is in the elliptical galaxy NGC 5128. It flared five times during a total observation time of 790 ks, yielding an approximate recurrent time of 1.8 days. The flares rise rapidly within 30 s by a factor of 200 to a peak luminosity of $8\times10^{39}$ erg s$^{-1}$, stay in a roughly steady ultra-luminous state for $\sim 200$ s, then decay over 4,000 s to the pre-flare level of $\sim 4\times 10^{37}$ erg s$^{-1}$. The optical counterpart is either a massive globular cluster, or a ultra-compact dwarf companion galaxy of NGC 5128 \citep{irwin16}.

The pre-flare and post-flare emission are found to be persistent during all the non-flare observation periods for the two sources. Absorbed power-law fits to the spectra of the persistent emission give photon indices $\Gamma \sim 1.6\pm0.5$. The in-flare spectra are more poorly constrained and can be fit either by an absorbed power law (photon indices $\Gamma \sim 1.3 - 1.6$) or by a disk blackbody ($kT \sim 1.3 - 2.2$ keV). No significant spectral evolution is found, either between the persistent and the flare periods, or during the flares.  

\citet{sivakoff05} reported two fast flares from an off-center source CXOU J124839.0--054750 in the elliptical galaxy NGC 4697, from two of five Chandra observations ($\sim 40$ ks each) of the galaxy. The flares have a peak luminosity of $\sim 6\times 10^{39}$ erg s$^{-1}$, a duration $\sim 70$ s and a count rate ratio of the flare to the persistent emission $\sim 90$. The photon counts are few (2-3 for each flare) so the statistical significance is not as good as those in \cite{irwin16}. The small number of photons likely causes the flare duration to be underestimated. In addition, \cite{jonker13} reported a bright fast flaring source in Chandra images of the old elliptical galaxy M86. Following \cite{irwin16}, we refer the four sources collectively as fast, ultra-luminous X-ray bursts (UXBs), and summarize them in Table \ref{tab:uxbs}. 

UXBs are unlikely of the same origins as those known bursting phenomena thought to happen to very young and highly magnetized neutron stars, such as Soft Gamma-ray Repeaters, Anomalous X-ray pulsars, or Types I and II X-ray Bursts, for the following reasons given by \cite{irwin16}. 1) UXBs mostly are found in old stellar population: globular clusters or compact dwarf companions of elliptical galaxies. 2) Compared with UXBs, the sporadic bursts from those previously known sources either last too short or have too low flare-to-pre-flare flux ratios. 3) The peak luminosities of UXBs would be super-Eddington for a neutron star. Note that although a handful of ultra-luminous X-ray pulsars have been found now which break this limit, they are usually thought to possess a high magnetic field strength ($\gtrsim 10^{12-13}$ Gauss) and probably a stable, copious mass supply (e.g., \cite{israel17}). These two conditions contradict either the oldness of the environment or the transient nature of UXBs.

If the peak luminosity of the flare is limited by the Eddington luminosity $L_{\rm Edd}$ of the black hole (BH)  (though one exception is a beamed emission), then the BH mass of $M \sim 10^{2-4}~ M_{\odot}$ is implied for most of the sources.  Each flare emits a total energy of $10^{42-43}$ erg for the two sources in \citet{irwin16}, which translates to a total accreted mass of $10^{-11\sim-10}~ \eta_{0.1}^{-1}~M_{\odot}$, where $\eta= \eta_{0.1}\times10\%$ is the radiative efficiency. Note that this mass estimate is a lower limit because the radiative luminosity is probably Eddington-limited so the efficiency could be much lower.

Under the condition of a pure black body, the loosely inferred temperatures could also hint at the size of the emission region: $R_{\rm BB}= [L/(4\pi \sigma T^4)]^{1/2}$. For source 1, $R_{\rm BB} \approx 5\times10^7$ cm, and for source 2, $R_{\rm BB} \approx 5\times10^6$ cm, implying very compact emission regions.  

If the flare represents a short period of accretion onto the IMBH during the periastron passage of a donor star on an eccentric or parabolic orbit, we argue that the flare's rise time corresponds to the duration during which the donor's tidally stripped mass returns to the pericenter. This duration is in turn equal to the other three time scales: the duration of the stripping, the sound crossing time of the donor, and the circularly orbital time of the transient disk formed at the pericenter radius. Only a white dwarf (WD) can have a sound crossing time as short as one minute.

We can not rule out the possibility that the binary comprises a stellar-mass BH and a WD, since the time scales that we will analyze in the next section are all independent of the BH mass. However, for a stellar-mass BH, the peak luminosity would be super-Eddington; to relax this limit one usually needs to invoke beaming of the radiation. 

In \S 2 we describe the tidal stripping and relevant physical time scales. We investigate the causes of the flare's flux decay and rise in \S 3 and 4, respectively, paying particular attention to the interaction of the stripped stream with a residual disk, and the subsequent accretion. We summarize and give further discussion in \S 5.  

\begin{table*}
\begin{center}
\small
\footnotesize
\caption{Observed fast ultra-luminous X-ray bursts as possible WD tidal stripping candidate events.}      \label{tab:uxbs}
\vspace{0.6cm}
\begin{tabular}{clcrrrccc}
\hline
Source & Host galaxy &  Number  & Rise & Decay & $L_X$ at peak & Factor of & Recurrence & Ref.\tablenotemark{d}\\
     &  &  of flares  &  time (s) & time (s) & (erg s$^{-1}$) & $L_X$ increase &  time (day)\tablenotemark{c} & \\
\hline \hline
1  & Ellip. NGC 4636  & 1 &  22  & 1,400  & $9\times10^{40}$  &  $\sim 100$ &  $>4$	& 1. \\
2  & Ellip. NGC 5128  & 5 &  30  &  4,000  & $8\times10^{39}$ & $\sim 200$  & $\sim 1.8$  & 1. \\
3  & Ellip. NGC 4697 &  2 &  $\lesssim 70$ & $\gtrsim 70$  & $6\times10^{39}$  & $\sim 90$ &	$\sim 1$ & 2. \\
4  &  Ellip. M86 (likely) & 1 & $\sim 20$ & $\sim 10^4$  & $6\times10^{42}$   & $\sim 600$ & $> 3.5$  & 3. \\
\hline
\end{tabular}
\end{center}
\tablenotetext{a}{For single-flare sources, the lower limit is the total observation time; for recurrent flaring sources, it is the total observation time divided by the number of flares, thus is a crude estimate.}
\tablenotetext{b}{1. \cite{irwin16}; 2. \cite{sivakoff05};  3. \cite{jonker13}.}
\end{table*}

\section{Periastron tidal stripping of the secondary}       \label{sec:stripping}

Consider an IMBH with a companion star orbiting around it on an elliptical orbit. Let $M$ and $M_*$ be the masses of the BH and the secondary star, respectively. The orbital period $P$ and the semi-major axis $a$ are related as $GM P^2 = 4 \pi^2 a^3$, thus, $a= 3\times10^{12}~M_3^{1/3}P_d^{2/3} ~\mbox{cm}= 10^4~(P_d/M_3)^{2/3}~R_S$, where $M_3= M/10^3 M_{\odot}$, $P_d= P/1$ day, and $R_S$ is the BH's Schwarzschild radius.

The secondary provides mass to the BH each time it moves to the pericenter whose distance from the BH is $R_p$, at which the secondary just fills its Roche lobe, i.e., the star's radius $R_*$ is about its Roche lobe size. Therefore \citep{paczynski71, eggleton83, sepinsky07}, 
\beq   \label{eq:rp}
R_p \simeq 2 R_* \left(\frac{M}{M_*} \right)^{1/3} \simeq 24~ R_*  \left(M_3  \frac{0.6 M_{\odot}}{M_*}\right)^{1/3}.
\eeq
In terms of the Schwarzschild radius, 
\beq    \label{eq:rp-rs}
R_p= 55~ M_3^{-2/3} \frac{R_*}{0.01 R_{\odot}} \left(\frac{0.6 M_{\odot}}{M_*}\right)^{1/3} R_S.
\eeq
Though here and after we normalize the secondary by typical numbers for a WD, the equations are valid for all types of stars.

The duration of the Roche lobe overflow (stripping) is $t_{\rm of} \simeq R_p/v_p$, where $v_p \simeq (2GM/R_p)^{1/2}$ is the secondary's orbital speed at $R_p$.  From equation (\ref{eq:rp}), it is easy to see that the duration of the stripping is roughly the internal dynamical time scale (also the sound crossing time) of the secondary star $t_{\rm dyn} \simeq (G \rho_*)^{-1/2} \simeq [R_*^3/(GM_*)]^{1/2}$, i.e., 
\beq     \label{eq:tof}
t_{\rm of} \simeq  2~ t_{\rm dyn} \simeq 6  \left(\frac{R_*}{0.01 R_{\odot}}\right)^{3/2} \left(\frac{0.6 M_{\odot}}{M_*}\right)^{1/2} \mbox{s}.
\eeq

After the periastron passage, since the stripped matter has a binding energy (with respect to the BH) of $E_{\rm min} = -GM R_*/R_p^2$, it follows an elliptical trajectory with a semimajor axis of $a_{\rm min}= 2 R_* (M/M_*)^{2/3}$, and a fallback time of  
\beq    \label{eq:tfb}
t_{\rm fb}= 2\pi \sqrt{\frac{a_{\rm min}^3}{GM}} \simeq 18 \left(\frac{M}{M_*}\right)^{1/2} t_{\rm dyn}.
\eeq
The above is valid under the assumption that the binding energy of the center of mass of the star is close to zero. This condition is satisfied as long as $a_{\rm min} \ll a$. Within the stripped mass, the spread of 
binding energy is small, therefore, the spread of $t_{\rm fb}$ is also small: $\delta t_{\rm fb} < t_{\rm of}$ (see Appendix).

The fourth time scale is the local circularly orbital time scale at $R_p$:
\beq		\label{eq:tcirrp}
t_{\rm cir}(R_p)= 2\pi \sqrt{\frac{R_p^3}{GM}} \simeq 18~ t_{\rm dyn}, 
\eeq

Thus, three of the above time scales are of the same order, $t_{\rm dyn} \sim t_{\rm of} \sim t_{\rm cir}(R_p)$. This makes $t_{\rm cir}(R_p)$ a unique quantity that depends only on the property of the donor star and not on the BH mass at all. For a sun-like star, $t_{\rm cir}(R_p)= 0.33$ day. For an evolved star such as the red giant Arcturus with $M_*= 1.1~M_{\odot}$ and $R_*= 25~R_{\odot}$, $t_{\rm cir}(R_p)= 40$ day.  

\begin{figure}[h]
\begin{center}
\includegraphics[width=8cm, angle=0]{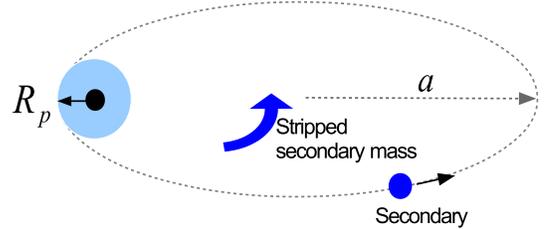}
\caption{This illustration shows that a secondary on an eccentric orbit about a black hole is tidally stripped at $R_p$. The stripped material falls back along a smaller orbit, hits the outer edge of a residual disk and replenish it with mass. The residual disk is left from previous stripping and replenishment.}
\end{center}
\end{figure}

\section{Cause of flux decay of bursts}  \label{sec:decay}

The duration of the mass supply at $R_p$ is approximately equal to the duration of the stripping, and again is approximately equal to the internal dynamical time scale of the donor, $t_{\rm dyn}$ (see Eqs. \ref{eq:rp}-\ref{eq:tof}). The closeness of numbers for $t_{\rm of}$ and $t_{\rm cir}(R_p)$ for a WD to the observed fast rise time ($\lesssim$ minute) of the flares already hints at a WD being the donor. But here let us consider the decay time of the flare firstly, regardless of this proposition. 

The disk's viscous time scale (representing the time that each mass element spends on its way to BH) at any radius $R$ is 
\beq	   \label{eq:tvis}
t_{\rm vis}(R)= \frac{t_{\rm cir}(R)}{2\pi \alpha} \left(\frac{H}{R}\right)^{-2},
\eeq
where $\alpha$ is the Shakura \& Sunyaev viscosity parameter, and $H/R$ is the disk hight-to-radius ratio. From Eqs. (\ref{eq:tcirrp}-\ref{eq:tvis}) one gets $t_{\rm vis}(R_p) \simeq 28~ \alpha_{0.1}^{-1} (H/R)^{-2}~ t_{\rm dyn}$, where $\alpha_{0.1}= \alpha/0.1$. This is the time scale over which the subsequent accretion rate (also the radiative luminosity) from a suddenly supplied mass at $R_p$ decays self-similarly (e.g., \cite{lynden-bell74}). 

If the disk is in the radiatively efficient, geometrically thin regime (Shakura \& Sunyaev disk), then $H/R \approx 0.02~(\alpha m)^{-1/10} \dot{m}^{1/5} r^{1/20}$. Here $m$, $\dot{m}$ and $r$ are BH mass, accretion rate and radius, normalized by $M_{\odot}$, $L_{\rm Edd}/(0.1 c^2)$ and $R_S$, respectively. In the advective cooling dominated, geometrically thick regime (slim disk), $H/R \approx 1$. The border line between the two regimes is $\dot{m} \sim r/10$ (e.g., \cite{kato98}). Since the accretion rate near the flare peak is around the Eddington rate, the real $H/R$ is likely between the two limiting values: $0.02 < H/R < 1$. 

Therefore, we see that  $t_{\rm vis}(R_p) \gg t_{\rm cir}(R_p) \sim t_{\rm dyn}$ since $H/R < 1$. This suggests that the flare decay time is more likely determined by $t_{\rm vis}(R_p)$, rather than by $t_{\rm dyn}$. The observed decay time of $10^3 - 10^4$ s means a rather short internal dynamical time scale of the donor $t_{\rm dyn} \sim 100~ \alpha_{0.1} (H/R)^2$ s. Therefore, a main-sequence donor is unlikely, since for instance, the Sun has $t_{\rm dyn} \approx 1.6\times10^3$ s; but the time scale is consistent with a WD being the donor.

\section{Causes of flux rise of bursts}

Now back to the rise time. Here we consider two independent scenarios. 

\subsection{Onset of accretion near ISCO}     \label{sec:onset}

The disk surface temperature typically drops with radius as $T(R) \propto R^{-p}$ where $p > 0$ (e.g., \cite{kato98}). Suppose $R_X$ is a radius in the disk within which the disk is hot enough to be X-ray bright. So the rise time corresponds to the time scale over which the ``head'' of the supplied mass accretes within the disk from $R_X$ to the BH, i.e., the viscous time scale at $R_X$: 
\beq      \label{eq:tx}
t_{\rm vis}(R_X)= 0.14~  \frac{M_3}{\alpha_{0.1}} \left(\frac{R_X}{R_S}\right)^{3/2} \left(\frac{H}{R}\right)^{-2} \mbox{s}.
\eeq
So if $R_X \approx 6.6~ R_S$ and $H/R \approx 0.2$, then $t_{\rm vis}(R_X) \sim 60$ s, consistent with the observed rise time. That is to say, most of the emission during the flare is radiated from close to the innermost stable circular orbit (ISCO) of the BH. 

\subsection{Stream-disk interaction}		\label{sec:stream}

After returning to the pericenter, the stripped material needs to dissipate its kinetic energy in order to circularize and form a disk. The specific energy to be dissipated is $\sim GM/R_p$. The dissipation is efficient when there is a residual accretion disk left from the previous episode of tidal stripping and mass replenishment. The existence of such a residual disk is supported by the fact that UXB sources show `persistent' emission before and after the flares, and during all the `non-flare' observation periods (see \cite{irwin16}). 

The outer radius of the freshly formed disk is $\sim 2 R_p$ if the bound material carries the same specific angular momentum as that of the star. Therefore, the returning stripped stream will collide with the outer disk at $R_p$, with a relative speed $\simeq 0.4 (GM/R_p)^{1/2}$ between the two. The portion of the disk mass that collides can be comparable to the stream's mass, while the total disk mass might easily exceed the latter because it is a cumulative residual from many previous rounds of mass replenishing and accretion.  The collision efficiently dissipates a large portion of the stream's orbital energy, with an equivalent efficiency (converting the rest-mass energy of the stream to heat) of $\eta \sim 0.001$ (cf. Eq. \ref{eq:rp-rs}). 

The dissipation heats the interaction site, whose size $R_s$ would be slightly larger than the width of the returning stream, but be smaller than the WD itself (because only its surface layer was stripped). A reasonable estimate would be $R_s \sim 10^7$ cm. This agrees with what was inferred from the spectral data of the two sources (see \S \ref{sec:intro}).   

The returning WD collides with the disk at $R_p$ as well. Because of the star's strong gravity and larger cross section (compared with the returning stream), this interaction probably scoops away a large chunk of the outer disk material through an extended bow shock in front of the moving star. These shock-heated material might produce a bright optical flare via Bremsstrahlung radiation.  

Once the stripped stream joins and replenishes the disk, the disk material drains into BH on the viscous time scale $t_{\rm vis}(R_p)$. The enhanced accretion rate (thus, the disk radiative luminosity) also subsides self-similarly on the same time scale (see \S \ref{sec:decay}). 

\section{Summary and discussion}
 
Four fast, ultra-luminous X-ray flaring sources (UXBs) have been discovered so far. We identify the flux decay of each flare with the viscous drainage of a suddenly supplied mass that was tidally stripped from a donor by a central IMBH ($\sim 10^{2-4}~ M_{\odot}$); the rapidness ($\sim$ hrs) of the decay suggests that the donor can only be a WD. The fast rise ($\sim$ minute) can be interpreted either as the onset of emission from the innermost region of the disk, or as due to the collision between the stripped stream and the outer disk when the former joins the latter at $R_p$.

The idea of a central IMBH is inspired by the observed peak luminosity of UXBs ($\sim 10^{40-42}$ erg s$^{-1}$). A stellar-mass BH ($\sim 10~M_{\odot}$) instead can not be ruled out but it has the issue of attaining super-Eddington luminosities.

The interval between two recurrent flares must be the eccentric orbital period $P$ of the donor. Independent from the type of the donor and the BH mass, it is straightforward to show that (see \S\ref{sec:stripping}) 
\beq 
1-e \equiv \frac{R_p}{a}= 2 \left(\frac{\pi t_{\rm of}}{P}\right)^{2/3}, 
\eeq 
where $e$ is the orbital eccentricity. If we identify the flare rise time with the duration of the stripping (\S\ref{sec:stream}), then for $t_{\rm of} \approx 1$ minute and $P \approx 1$ day, we get $e \approx 0.97$. However, if instead the flux rise corresponds to the onset of X-ray emission from the innermost region of the disk (\S\ref{sec:onset}), then $t_{\rm of}$ can be longer and the above constraint on $e$ relaxes. Subsequent X-ray monitoring of those UXB sources would be key to constraining the binary parameters and verifying the tidal stripping scenario .  

The source reported by \cite{sivakoff05} has no clear optical counterpart and the nearest globular cluster (GC) is 1\arcsec.8 ($\gtrsim$ 1 pc) away. It is possible that this source was ejected from its parent GC at the typical escape speed from a GC (e.g., $\sim$ 10 km s$^{-1}$) $\sim$ 1 Myr ago. The ejection of an IMBH can be done by the gravitational recoil resulting from a merger with a stellar-mass BH (e.g., \cite{favata04,fragione18}). It acquires a WD on its way out either by tidal capture of a single red giant (e.g., \cite{fabian75,kalogera04}) or by exchange interaction with a binary containing a WD (e.g., \cite{ivanova10}). Hierarchical triple interaction via the Kozai mechanism is needed to maintain a high eccentricity of the IMBH-WD binary. This triple system probably forms when the latter encounters another ordinary binary before it escapes the GC (e.g., \cite{ivanova10}).

The optical counterpart to the \cite{jonker13} source in the M86 galaxy is very faint, with an absolute magnitude $M_{i'} > 25.6$, roughly corresponding to a bolometric luminosity of $< 10^4 L_{\odot}$. The galaxy is falling into the Virgo cluster and its gas is being stripped away into a long stream. Moreover, M86 shows signs of a recent wet minor merger with another galaxy SDSS J122541.29+130251.2 whose stars are being ripped off and follow the gas stream. The projected location of the UXB source lies close to the gas and star stream \citep{jonker13}. So it is very likely that the parent GC of this UXB was tidally disrupted or stripped during the wet merger  (e.g., \cite{kruijssen12}), leaving its low mass core (which hosts the UXB) ejected in the direction of the stream.

Tidal disruptions of WDs by IMBHs have been widely studied \citep{krolik11, macleod14, macleod16, law17}, which differ from the tidal stripping considered here in several ways. First, for parabolic orbits, the event rate of stripping is slightly higher than the disruptions since $R_p \approx 2 R_t$. Second, for elliptical orbits, the system spends several orders of magnitude longer in time during which tidal stripping repeatedly occurs than the time when a disruption occurs (see below). 

Third and the most importantly, the radiation time scales from a WD disruption event shall be much longer. In such a event, the fallback mass rate is extremely super-Eddington ($\sim 10^7\times$ at peak) and it remains above the Eddington rate for $\sim 1$ year. One would expect enormous mass ejection in forms of quasi-spherical outflow during this long period due to energy dissipation from debris stream collision and (later) central accretion. Any high-energy emission from near the BH would be reprocessed by the opaque outflow and the photospheric emission of the latter dominates the observation, similar to a main-sequence stellar TDE by IMBHs \citep{chen18}. Therefore, the emission of a WD TDE will be at lower photon energies (UV to soft X-rays) and last much longer (months to a year). Although the fallback time of the most bound debris (which falls back earliest) is relatively short ($\sim$ 10 minutes), it is still very hard to envisage a fast transient could emerge which brightens to $L_{\rm edd}$, shines for 1 minute and then decays. Moreover, the currently detected UXBs lack any sign of significant spectral or absorption evolution, which disfavors a scenario in which the flare had appeared in the earliest minute of a WD TDE and it was quickly obscured by the launching of an outflow.              

\cite{zalamea10} studied the impact of gravitational wave (GW) emission on the orbit of a WD-IMBH ($10^5~M_{\odot}$) binary which experiences tidal stripping. They show that the slow decrease of $R_p$ (thus, the increase of stripped mass $\delta M$ in each passage) experiences two stages. Stage (i) is controlled by the gentle GW emission. The fractional change of $R_p$ during each orbit is $\gamma \equiv -P \dot{R}_p/R_p \sim 10^{-5} M_5^{2/3 }$. Stage (ii) is controlled by the WD mass loss, hence, the increase of $R_*$. Stage (i) lasts longer (roughly $\gamma^{-3/5} \sim 10^3$ orbits) and mass loss is gentle, but the mass loss accelerates in stage (ii) until it reaches total disruption ($\sim 200$ orbits). To scale down to $M= 10^3 M_{\odot}$, stage (i) would be even longer, $\sim 10^4$ orbits. Due to its dual-signal nature, such systems are interesting targets for next-generation GW detectors, e.g., DECIGO and Einstein Telescope, with the aid of long-term X-ray monitoring.

\acknowledgments

The author is grateful to Pawan Kumar, Xin-Wen Shu, Wei-Min Gu, Song Wang and Ning Jiang for enlightening discussion and comments, particularly to Chris Matzner for reading carefully through the manuscript, and thanks the anonymous referee for suggestions that improved the quality of the manuscript. This work is supported by NSFC grant 11673078.

\begin{appendix}
\setcounter{figure}{0} 
\renewcommand{\thefigure}{A\arabic{figure}}
\setcounter{equation}{0} 
\renewcommand{\theequation}{A\arabic{equation}}

\begin{figure}[h]
\begin{center}
\includegraphics[width=8cm, angle=0]{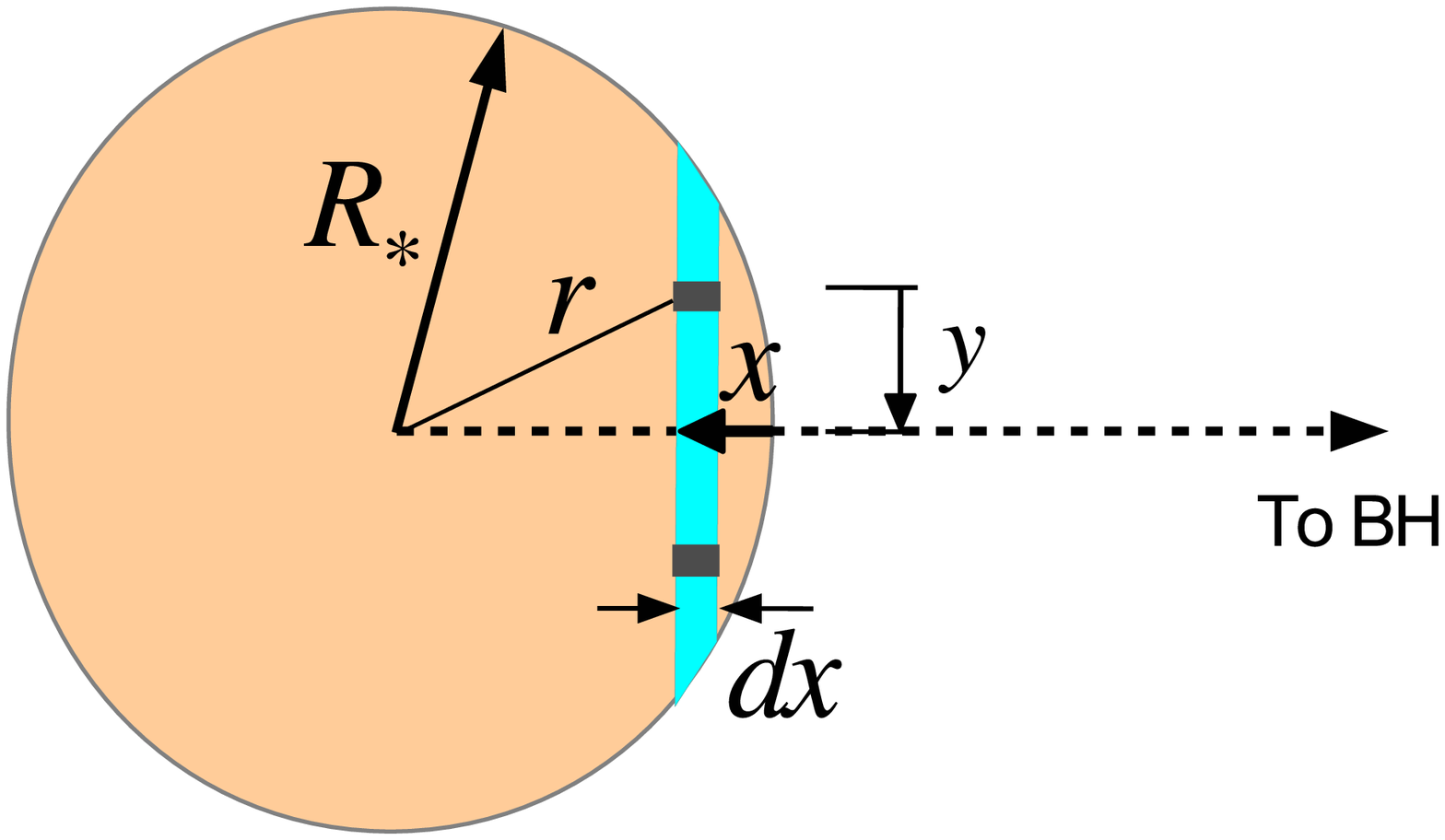}
\caption{The white dwarf donor at the pericenter when its surface layer of a depth $x$ facing the black hole is tidally stripped. Here we neglect tidal deformation.}   \label{fig:stripping}
\end{center}
\end{figure}

Here we take a closer look at the tidal stripping, in order to characterize the mass, depth, and specific energy distribution of the stripped layer. Let $x \ll R_*$ be the depth of the stripped surface layer, as is illustrated in Figure \ref{fig:stripping}. Consider a ring of radius $y$ within the layer at depth $x$. Each point on the right is at distance $r=\sqrt{y^2+(R_*-x)^2}$ from the center of the star. At any fixed depth $x$, the mass per unit depth is 
\beq
\frac{dM}{dx}= 2\pi \int_{0}^{\sqrt{R_*^2-(R_*-x)^2}} \rho(x,y) y dy.  
\eeq 
Changing the variable from $y$ to $r$ by $ydy= rdr$, the integral becomes 
\beq
\frac{dM}{dx}= 2\pi \int_{R_*-x}^{R_*} \rho(r) r dr.  
\eeq
Since $x \ll R_*$, we can approximate $r \approx R_*$ and move it out of the integrand. Then we use another depth variable $x'= R_*-r$ to rewrite the above to
\beq
\frac{dM}{dx}= 2\pi R_* \int_{0}^{x} \rho(x') dx'.
\eeq

For simplicity, we assume the surface structure of the white dwarf is described by a polytrope of $P= K \rho^{\Gamma}$ with $\Gamma= 5/3$, the same as in the deeper region where electrons are degenerate and non-relativistic. Once $M_*$ and $R_*$ are given, the value of $K(M_*, R_*)$ is known from numerically solving the Lane-Emden equation. The hydrostatic equilibrium at the surface $dP/dx= GM_* \rho /R_*^2$ gives the density structure there
\beq
\rho(x)= A \bar{\rho} \left(\frac{x}{R_*}\right)^{3/2},
\eeq  
where $A \simeq 3.8$ and $\bar{\rho}= M_*/(4\pi R_*^3/3)$ is the average density. Therefore,
\beq
\frac{dM}{dx}= \frac{3}{5} A \frac{M_*}{R_*} \left(\frac{x}{R_*}\right)^{5/2}.
\eeq
The fraction of the total stripped mass is
\beq    \label{eq:deltaM}
\frac{\delta M}{M_*}= \frac{6}{35} A \left(\frac{x}{R_*}\right)^{7/2}.
\eeq

Equation (\ref{eq:deltaM}) provides a relation between $\delta M$ and $x$, so that one could estimate the depth ratio $x/R_*$ from $\delta M/ M$. The observed fluence of the two UXBs in \cite{irwin16} suggest about $\sim 10^{-10}~M_{\odot}$ of mass is accreted in each case. This implies the depth of the stripped layer is $x/R_* \sim 10^{-3}$.  An alternative version of equation (\ref{eq:deltaM}) is $\delta M / M_* \propto (x/R_*)^{5/2}$ as was given by \citet{zalamea10} who adopted a spherical-shell shape of the stripped layer. There, the stripped layer depth ratio is even smaller, $x/R_* \sim 10^{-5}$, for the same $\delta M / M_*$. 

Within the stripped mass, the spread of the binding energy relative to the BH is very small, $\delta E /E \simeq x/R_*$, which means a very small spread of the returning time $\delta t_{\rm fb} / t_{\rm fb} \simeq 3\delta E /(2 E) \simeq 3x/(2R_*) \ll 1$. Since $t_{\rm fb} \sim 9(M/M_*)^{1/2} t_{\rm of}$ (Eq. \ref{eq:tfb}), then that means a small ratio of the spread of returning time over the ``length'' of the stream, $\delta t_{\rm fb} / t_{\rm of} \sim 14(M/M_*)^{1/2} x/R_* \sim 0.6$ for $M/M_*=10^{3.2}$ and $x/R_*= 10^{-3}$. Therefore, the duration over which the stream of the stripped matter returns to the pericenter and collides with the residual disk is set by the duration of stripping $t_{\rm of}$.

\end{appendix}

\end{CJK*}
\end{document}